\documentclass[]{spie}  
\usepackage[]{graphicx}
\usepackage{times}
\usepackage{amsmath}
\usepackage{float}
\usepackage{subfig}
\usepackage{multirow}
\usepackage{multicol}

\pdfoutput=1

\newcommand{\desspot}{DeSSpOt }
\newcommand{\desspots}{DeSSpOtS }

\title{DeSSpOt : an instrument  for stellar spin orientation determination}
\author{Anna-Lea Lesage, Magnus Schneide, G\"unter Wiedemann
\skiplinehalf
\supit{}Hamburger Sternwarte, Gojenbergsweg 112, 21029 Hamburg
\authorinfo{(Send correspondence to A-L Lesage, e-mail: alesage@hs.uni-hamburg.de)}
}
\begin{document}
\maketitle

\begin{abstract}

We designed and constructed a special instrument to enable the determination of the stellar's spin orientation. The \textbf{D}ifferential imag\textbf{e} rotator for \textbf{S}tellar \textbf{Sp}in \textbf{O}rien\textbf{t}ation, \textbf{DeSSpOt}, allows the simultaneous observations of two anti-parallel orientations of the star on the spectrum. 
On a  high resolution \'echelle spectrum, the stellar rotation causes a slight line tilt visible in the spatial direction which is comparable to a rotation curve. We developed a new method, which exploits the variations in these tilts, to estimate the absolute position angle of the rotation axis. The line tilt is retrieved by a spectro-astrometric extraction of the spectrum. \\
In order to validate the method, we observed spectroscopic binaries with known orbital parameters. The determination of the orbital position angle is equivalent to the determination of the stellar position angle, but is easier to to detect. \\
\desspot was successfully implemented on the high resolution Coud\'e spectrograph of the Th\"uringer Landessternwarte Tautenburg. The observations of Capella led to the determination of the orbital position angle. Our value of $37.2^\circ$ is in agreement with the values previously found in the literature. As such we verified that both method and instrument are valid.\\

\end{abstract}

\section{INTRODUCTION}

Orientation of the rotation axis is unknown for most stars. Absolute position and inclination angles have been measured only for a handful of stars. The first method proposed for the  determination of these parameters relied on Differential Speckle Interferometry \cite{Beckers}, and was successfully applied to Aldebaran in 1995 \cite{Lagarde}. However the observations, made with a single aperture, lacked in signal to noise due to the short exposure, which made this technique effective only for the brightest stars. Differential Speckle interferometry evolved in the last decade into long baseline interferometry. The improvements both in instrumentation and theory have made it possible to image the shape of some stars. The measurements of the oblateness for Altair \cite{Monnier}, Achenar \cite{deSouza}, Vega \cite{Peterson} or Alderamin \cite{Zhao} permitted the determination of inclination and position angle of the spin axis. However, interferometric observations require to simultaneous use of several telescopes.

Spectro-astrometry is an alternative approach for probing structures in the milliarsecond scale, i.e. below the diffraction limit of the telescope. The method relies on the conservation of the spatial information along the slit direction, and can be used with any long slit or \'echelle-spectrograph. By measuring the wavelength dependence of the position with respect to the photocentre along the spectrum, any asymmetries in the spectral energy distribution of the source can be followed down to the milliarcsecond scale on the resulting position spectrum. The latter outlines structures which are displaced from the continuum along this one direction. In order to constrain their shape or position, it is necessary to probe several directions thus to take several exposures with different slit orientations, or to rotate the image of the source on the slit. The method has been already successfully applied to the determination of jets\cite{Whelan}, separation of binaries\cite{Bailey}, and position of stellar spots\cite{Voigt}.

In this paper we explain in \S2 how a spectro-astrometric analysis of an \'echelle-spectrum from a star permits the determination of the position angle. Then we describe in \S3 the adequate instrumentation \desspot and the first light results. Finally, in \S4 we conclude with the realisation of a DeSSpOt-spectrograph for an use at Hamburg Observatory.

\section{THE METHOD}

\begin{figure}[H]
 \centering
  \subfloat[Notations]{\label{fig:convention}\includegraphics[width = 0.3 \textwidth]{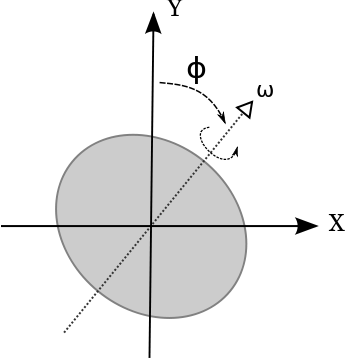}}
 \hspace{1em}
  \subfloat[2D broadening function]{\label{fig:d2dop}\includegraphics[width = 0.6 \textwidth]{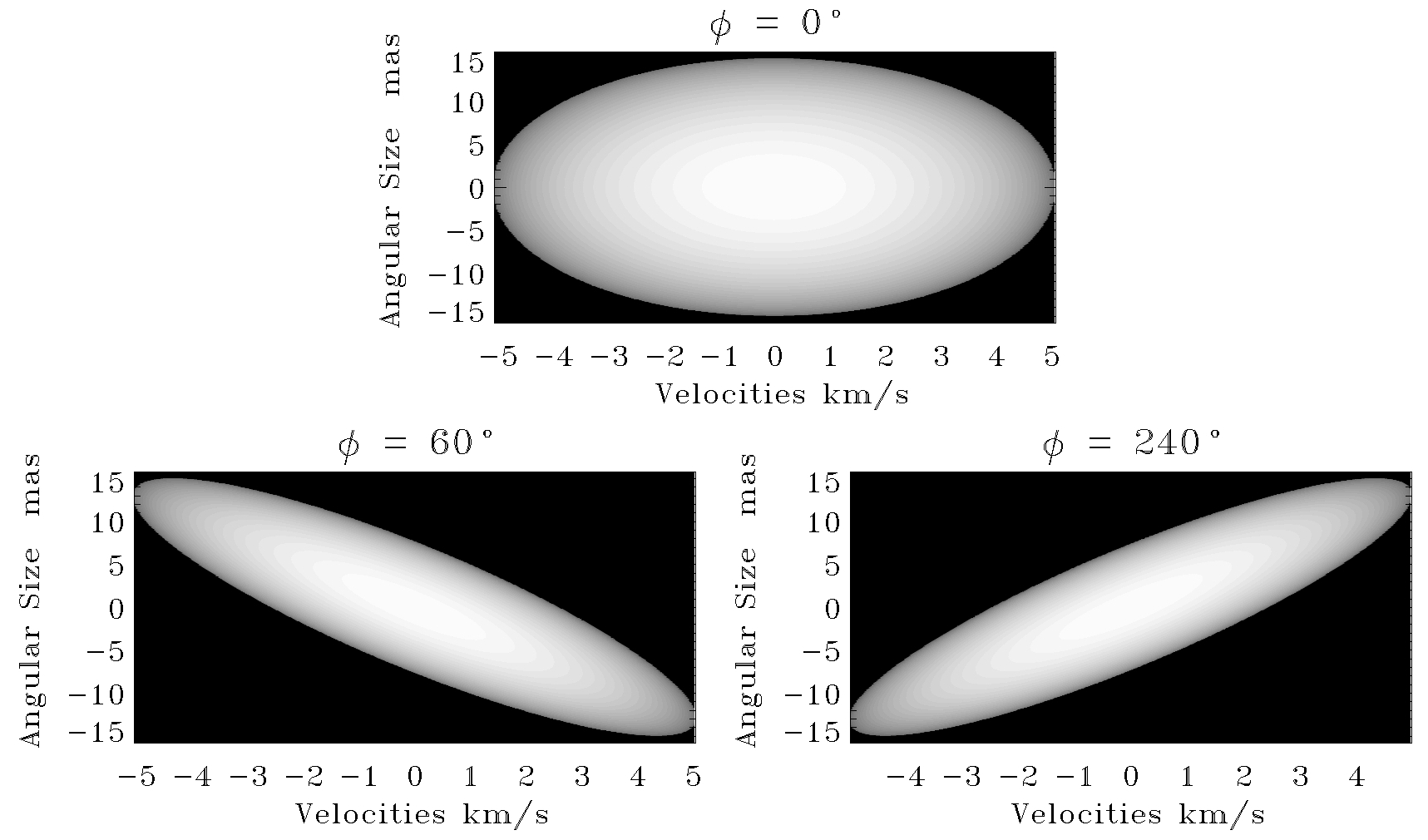}}
   \caption{\emph{Left}: Conventions adopted in this paper: X,Y are the slit coordinates along the spectral and spatial direction respectively. $\phi$ is the relative angle between the slit spatial axis and the stellar rotation axis. \emph{Right}: The 2D broadening function calculated, for a star with 30 mas angular size, from the total wavelength shift for each point (X,Y) and implemented into the Limb-darkening function $ I_c/I_o = 1-\epsilon +\epsilon\cos\theta$. }
\end{figure}

In this section we use the conventions of Fig \ref{fig:convention}: the star rotates counter-clockwise, it is projected on the slit, and $\phi$ is the relative angle between the slit's spatial axis and the star's spin axis. 
Furthermore, the star is no longer assumed to be a point source but has an extended size  below the diffraction limit of the telescope. As a result, the effects of the stellar rotation are stretched in the spatial and spectral directions of the spectrum. The two dimensional Doppler broadening function, which takes into account the stellar's spin orientation and the stellar size on the slit, has the form of an elongated and slanted ellipse as illustrated on Fig \ref{fig:d2dop}. The semi major axis is the wavelength shift, expressed here in velocities, caused by the stellar rotation, while the semi minor axis is the apparent angular size of the star $R_{star}$. The tilt  angle $\gamma$ of the ellipse is given by: 
\begin{equation}
 \gamma =  \arctan \left ( \frac{V_{rot} \lambda \sin i  \sin \phi }{c R_{star}} \right )
\end{equation}
where $V_{rot}$ is the stellar rotational velocity, and $\lambda$ the wavelength. In the observed spectrum all the lines are tilted. However the seeing, acting like a blurring function, degrades the spatial information. Consequently, the larger the star, in other words the larger the semi minor axis of the ellipse, the clearer the line tilt remains in the spectrum. Therefore, the best stellar candidates have large apparent diameter and many narrow absorption lines. 

\begin{figure}
 \centering
  \includegraphics[width= 0.85 \textwidth]{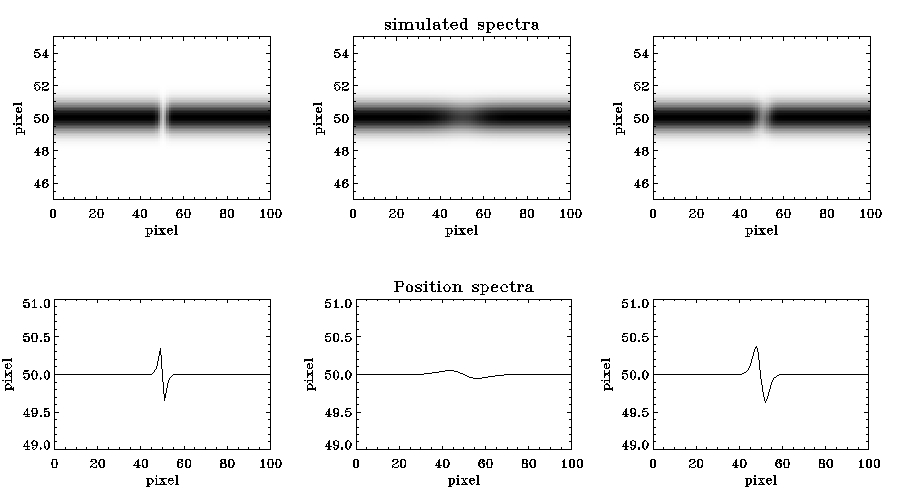}
  \caption{\emph{Top}: Simulation of a tilted line from the convolution of the 2D-broadening function with a Gaussian-like absorption line. On the top-left side, the line is very deep and narrow but tilted by 0.5 pixel; on the top middle, the line is rotation broadened and tilted by 2 pixel, finally on the right top side the line is slightly broadened and tilted by 2 pixel. The tilts are exaggerated here for understanding purpose. 
  \emph{Bottom}: The spectro-astrometric signature of the line tilt resulting from the position of the centroid of the order for each pixel. The signal is more pronounced for slow rotating stars (bottom left and bottom right cases) than for fast rotators (bottom middle case).}
  \label{fig:signature}
\end{figure}

The spectro-astrometric signature, obtained by fitting the barycentre of the order at each pixel, is a shift from the continuum on the position spectrum correlated with the line tilt as illustrated in Fig \ref{fig:signature}. 
While the shape of the signal remains similar, the amplitude of the signal is correlated to the contrast of the line. In addition, the signal is a function of sin$\phi$ and is inverted for $\phi$ greater than 180$^\circ$ when the line's inclination changes. By monitoring the amplitude variation of the signal and its inversion point, we deduce the position angle of the star. In order to do so, the star has to be observed under different orientations to probe different angles $\phi$. 

In order to identify in the position spectrum the seeing component from the stellar component in the signal, we are compelled to capture simultaneously two spectra with anti-parallel orientations. Accordingly, we developed an instrument, the \textbf{D}ifferential imag\textbf{e} rotator for \textbf{S}tellar \textbf{Sp}in \textbf{O}rien\textbf{t}ation, hereafter \textbf{DeSSpOt}, to fulfil this requirement. \\
The expected amplitude of the signal is in the sub-pixel range, between 1-5\% of a pixel depending on the stellar parameters and the spectrograph's characteristics. At this scale, the detection is sensitive to every external error source. We identify the two possible main error sources in the position spectrum which could influence the signal: 
\begin{itemize}
 \item Astigmatism in the spectrograph and instrumental artefacts. The former causes a signal for all lines in the position spectrum equivalent to the rotation signal. On the opposite,  artefacts generate local distortions. Both are assumed to vary on large time-scale, and to remain constant during an observation run. As a result their signals in the position spectrum is also constant during the night, and can be cancelled out by subtracting position spectra of anti-parallel orientations. 
  \item Seeing leads to rapid changes in the point spread function of the star on the slit. It disturbs the line shape or even induce false signal on the position spectrum. Due to the short time-scale of seeing, typically around 10 ms, two consecutive exposures are no longer directly comparable. As a result the subtraction of anti-parallel position spectra is ineffective as long as the images are taken in succession. 
\end{itemize}

\section{DeSSpOt}

\subsection{Instrumental requirements}

\desspot shall be inserted on existing spectrographs. This results in a set of constraints on the instrument for a  smooth use of both \desspot and spectrograph.
We identify three types of requirements and constraints on the instrument. These types are respectively of geometrical, optical, and mechanical nature. We resume in Table \ref{tab:requirements} the main requirements and constraints for the instrument. 

The optical axis and the aperture ratio of the telescope must be preserved at \desspot's output. By keeping the optical axis, the exiting beam hits both slit and spectrograph's collimator. The conservation of the aperture ratio assures a perfect illumination of the collimator, thus the full performance of the spectrograph while limiting light losses. 
The grating of the spectrograph is fully used for high resolution spectra. Therefore, the two beams from \desspot are in the same pupil field. The separation and recombination of the beams has to fulfil this constraint and also limit the light losses and the global size.
The insertion of the instrument shall also not displace the position of the pupil field configuration inside the spectrograph.

\begin{table}
 \centering
  \begin{tabular}{|c|p{10cm}|c|}
   \hline
   Type & Requirement or constraint & Number \\ 
    \hline \hline
   \multirow{2}{*}{Geometrical} & Minimal size & 1 \\
    & Available space on the TLS Spectrograph : 120 mm in the optical axis,  200 mm laterally & \\
  \hline
   \multirow{5}{*}{Optical} & Keep the optical axis of the telescope & 2 \\
     & Preserve the F ratio and focus position of the telescope & 3 \\
    & Limit the light losses & 4 \\
    & Rotate the beam by $180^\circ$ & 5 \\
    &Sufficient separation of the orders on the spectrograph & 6 \\
  \hline
   \multirow{2}{*}{Mechanical} & Adjustable mounts & 7 \\
    & Stable for minimal vibrations & 8 \\
   
    \hline
  \end{tabular}
\caption{List of the requirements and constraints defined for \desspot.}
\label{tab:requirements}
\end{table}

The geometrical constraint, which is the  most critical of the instrument, and the need to rotate the beam by 
$180^\circ$, have lead us to search for an alternative to the multiple mirrors solution\cite{Denisov}. Indeed the latter limits the light losses, but it requires at least three mirrors which have to be adjusted very precisely. Therefore, it does not fulfil the size and mechanical requirements. Contrary to this are Dove prisms, which have an increasing use in astronomy as natural beam rotator. They rotate the image by twice their own rotation angle.

\subsection{Optical layout and concept}

The function of DeSSpOt is to image two anti-parallel orientations of the star on the slit. Fibre-fed spectrographs are not suited  because they do not conserve the spatial information. Hence, \desspot has to be inserted between the telescope output and the entrance of the spectrograph. The concept is to separate the incoming light from the telescope into two beams of which one is rotated by $180^\circ$, and to  recombine them to focus on the slit. Two spectra are captured by the detector during one exposure. 

\begin{figure}[h]
 \centering
\includegraphics[width = 0.5 \textwidth]{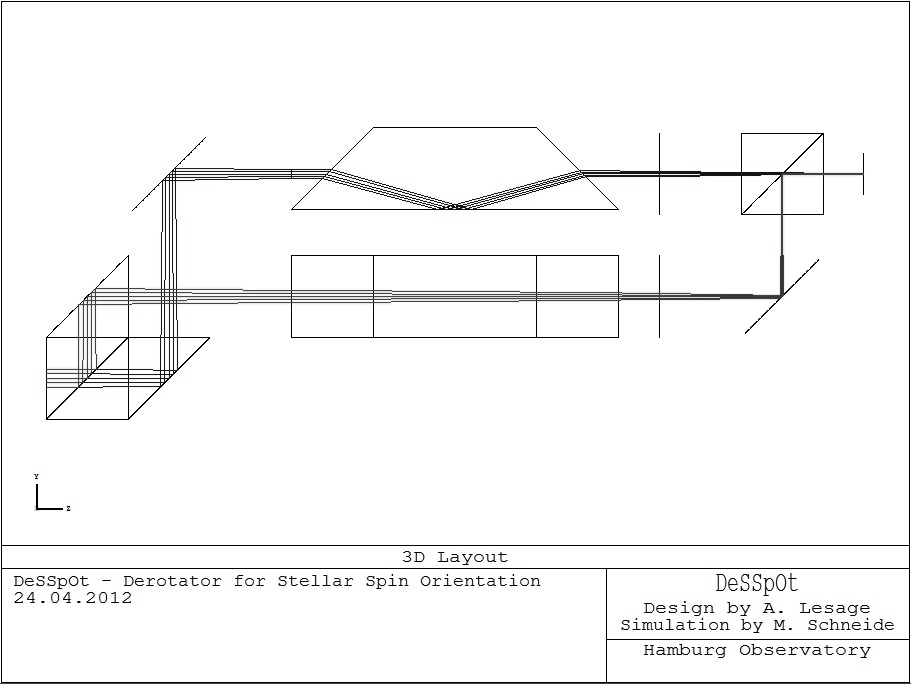}
\caption{Optical layout of DeSSpOt: the light from the telescope is coming from the left with a F/46 aperture ratio, after passing the instrument the beams are focused on the spectrograph's slit.}
\label{fig:opticdesign}
\end{figure}

The instrument is built after the layout in Fig \ref{fig:opticdesign}. The incoming light from the telescope is separated according to linear polarization. Two small right corner prisms are cemented on the output faces of the beamsplitter. They redirect the light perpendicularly to the output direction. One beam is deviated by $90^\circ$ with an adjustable aluminium mirror. Each beam is then sent towards one of the Dove prism, which are rotated respectively by $180^\circ$ and $90^\circ$, in order to keep an identical optical path lengths for both channels.  
The size of the Dove prisms is kept as small as allowed by the beam diameter. Finally the beams are directed towards a second polarisation beamsplitter for recombination. The total path length is less than 200 mm and would barely influence the position of the pupil plane in most spectrographs.

The separation and recombination of the beams by polarisation is preferred to an intensity separation. The recombination in intensity mode would either result in excessive light losses in one channel or in additional optics to fully use the light of both channels. In addition, stellar light being mainly unpolarised, the intensity of the s- and p-polarised beams are similar as long as the telescope does not induce itself large polarisation. \\
Adjustable mirror mounts correct for the deviations of the prisms, and perform the alignment of the beam on the slit. 

The Dove prisms are optimised for collimated beams. However the image quality is barely affected by using slightly converging beams, e.g. a F/46 aperture ratio.
The Dove prisms are made of $CaF_2$ with a base angle of $45^\circ$ and a side height of 10 mm. The length of the Dove prism is dictated by its side height, the base angle and the material \cite{Sarel}. The lengths calculated for the previous prism's size at different materials are presented on Table \ref{tab:materials}. Finally the material choice is the outcome of the error analysis of the prism described in the following section. 

\begin{table}
 \centering
 \begin{tabular}{cc|c|c|c|}
  \cline{3-5}
  & & \multicolumn{3}{|c|}{Materials} \\
\hline \hline
 \multicolumn{2}{|c|}{Properties}  & N-BK7 & $CaF_2 $ & LiF \\ 
\hline
 \multicolumn{2}{|c|}{Refractive index at $\lambda$ = 550 nm} & 1.5185 & 1.4348 & 1.3930 \\
\hline
  \multicolumn{2}{|c|}{Length (D=10 mm, $\alpha=45^\circ$)}  & 42.21 mm & 46.12 mm & 48.68 mm\\
\hline
  \multicolumn{1}{|c|}{\multirow{2}{*}{Transmission (s-polarization)}}& $0^\circ$  & 80.03 \% & 80.86\% &  80.99\% \\
 \multicolumn{1}{|c|}{} & $90^\circ$ & 79.31 \% & 83.54 \% & 85.47\% \\
\hline
 \end{tabular}
\caption{Properties of Dove prisms in the above materials. The transmission rates are similar for p-polarised light. They are calculated with the optical analysis programme ZEMAX. }
\label{tab:materials}
\end{table}

\subsection{Analysis of the Dove prisms}

The Dove prisms are the key elements of the \desspot instrument. Effects of manufacturing errors are investigated here. We analyse the characteristics of the prisms for three potential materials with decreasing refractive index and Abbe number: N-BK7, $CaF_2$ and LiF.

\begin{figure}[H]
 \centering
  \subfloat[Vertical deviation from base angle errors]{\label{fig:coltolerance} \includegraphics[width = 0.5 \textwidth]{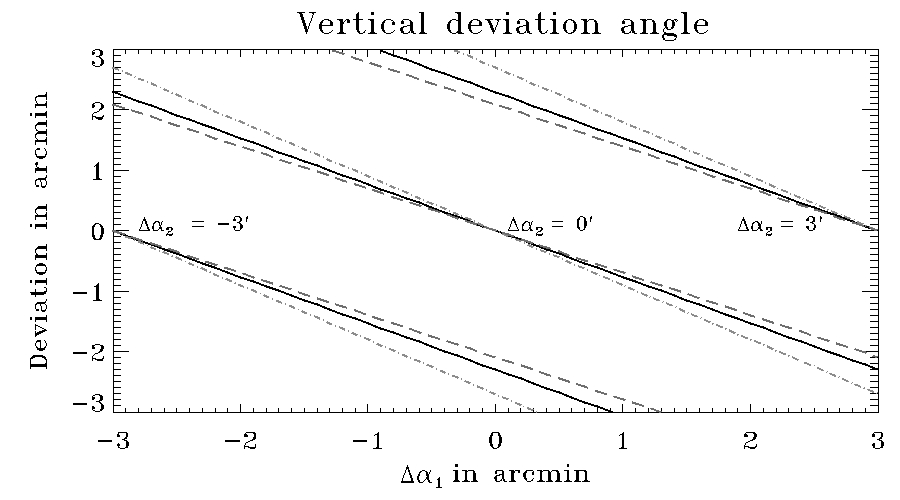}}
  \subfloat[Lateral deviation from pyramidal errors]{\label{fig:pyramidal} \includegraphics[width = 0.5 \textwidth]{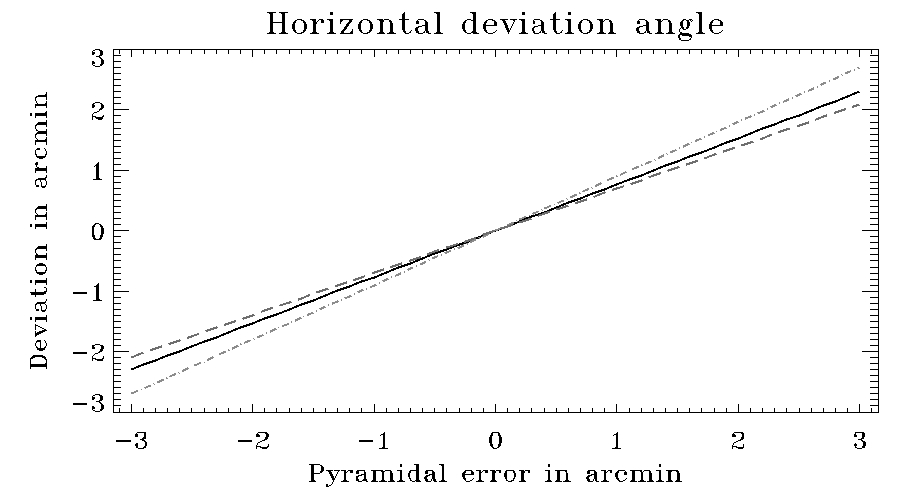}}
  \caption{Two types of beam deviations caused by manufacturing errors for $CaF_2$ in black straight lines, N-BK7 in dash-dotted lines, and LiF in dotted lines.}
\end{figure}

The tolerances for the base angles are at the most 3 arcmin. Small divergences in the value of the base angles lead to a vertical deviation of the beam. Similarly, the pyramidal error\cite{Strojnik}, resulting from non-orthogonality between the reflecting and the side plane of the prism, causes a horizontal deviation of the beam. The global deviation is then defined by  \begin{math} \delta_{tot}^2 = \delta_{V}^2 + \delta_{H}^2 \end{math}, where $\delta_V$ and $\delta_H$ are the vertical and horizontal deviations. As illustrated in Fig \ref{fig:coltolerance} and Fig \ref{fig:pyramidal}, these values remain small enough to be corrected with adjustable mirrors. The highest and lowest deviations are reached for N-BK7 and LiF prisms respectively.

\begin{figure}[H]
 \centering
  \subfloat[Prism diffraction]{\label{fig:dovdiff} \includegraphics[width = 0.5 \textwidth]{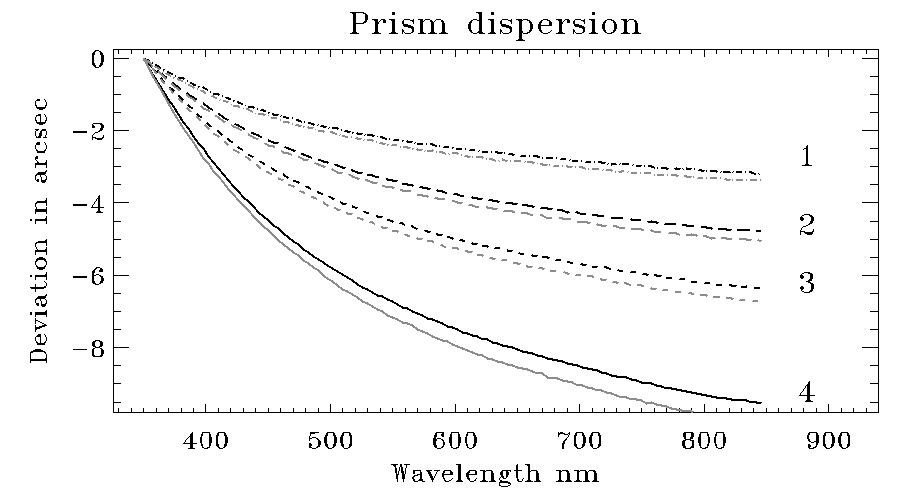}}
  \subfloat[Atmospheric diffraction]{\label{fig:atmdiff} \includegraphics[width = 0.5 \textwidth]{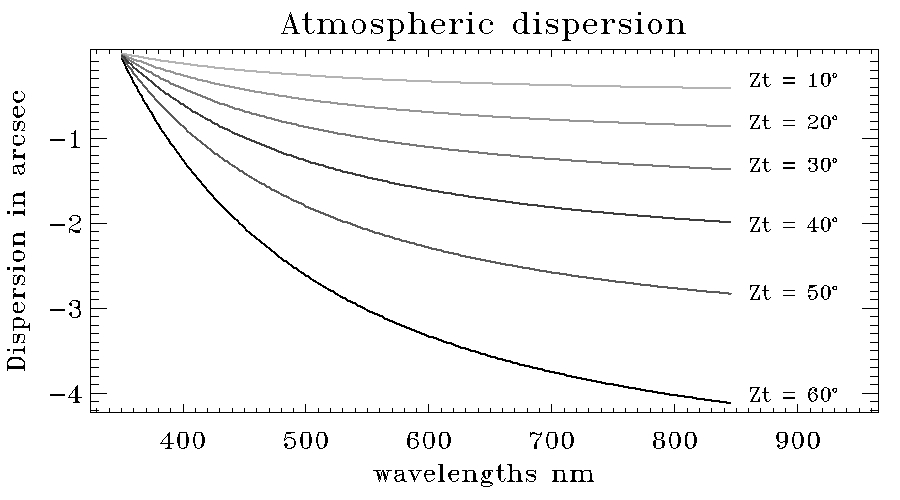}}
  \caption{\emph{Left}: The diffraction of the prism for four error cases: \textbf{1}, $\Delta \alpha_1 - \Delta \alpha_2$ = 2 arcmin; \textbf{2}, $\Delta \alpha_1 - \Delta \alpha_2$ = 3 arcmin; \textbf{3, }$\Delta \alpha_1 - \Delta \alpha_2$ = 4 arcmin; \textbf{4},  $\Delta \alpha_1 - \Delta \alpha_2$ = 6 arcmin. In black, calculated for collimated light, in grey for a F/13 aperture ratio input beam. The diffraction is more
 pronounced for uncollimated light. \emph{Right}: Atmospheric diffraction, for several zenith angles Zt,
 calculated from the Ciddor Model \cite{roe}.}
\end{figure}

In addition, we investigate how the prism tolerances influence the diffraction of the output beam. The dispersion angle is more pronounced for large base angle errors, e.g. the increasing errors plotted in Fig \ref{fig:dovdiff}. Moreover, the dispersion angle is in the same order of magnitude than the atmospheric dispersion angle, see Fig \ref{fig:atmdiff}. Dispersion of the beam in the spatial direction causes a shift in the position of the continuum in the position spectrum. On the contrary, dispersion in spectral direction leads to variations in the incoming angle of the grating, which in turn cause changes in the wavelength solution of the spectrum. The latter effect is minimal and does not influence the resolution of the spectra. 
 
\subsection{Installation on the TLS-spectrograph}

\subsubsection{Spectrograph and prototype description}

The Coud\'e-spectrograph of the Th\"uringer Landessternwarte, hereafter the TLS-spectrograph, is a high resolution \'echelle-spectrograph with peak resolution of 67 000 in the visible. It is mainly used for RV exoplanet search. It is built after a white pupil design. The light from the telescope, with an aperture ratio of F/46, is collimated to a beam with 150 mm diameter. It is diffracted by an \'echelle-grating, with a blaze angle of $65^\circ$. The  cross dispersion is carried out by a grism. There are three standard grisms which provide, respectively, a coverage in the blue, in the visible and in the near infrared. Finally the spectrum is formed on a 2k x 2k CCD detector. The two pixel resolution of 67 000 is reached with a slit width of 0.52 arcsec. In the visible channel, the distance between two adjacent orders is at least 34 pixel with a limited slit length. In order to avoid an overlap of the orders in the red when observing with DeSSpOt, we used an additional grism to assure a larger separation of the orders.  

Its maximum efficiency is reached at the blaze wavelength of 560 nm. The resulting inter-order separation was increased to 53 pixel with full prism length. The new wavelength coverage extended from 420 nm to 650 nm.

\begin{figure}
 \centering
  \subfloat[\desspot prototype]{\includegraphics[width = 0.4 \textwidth]{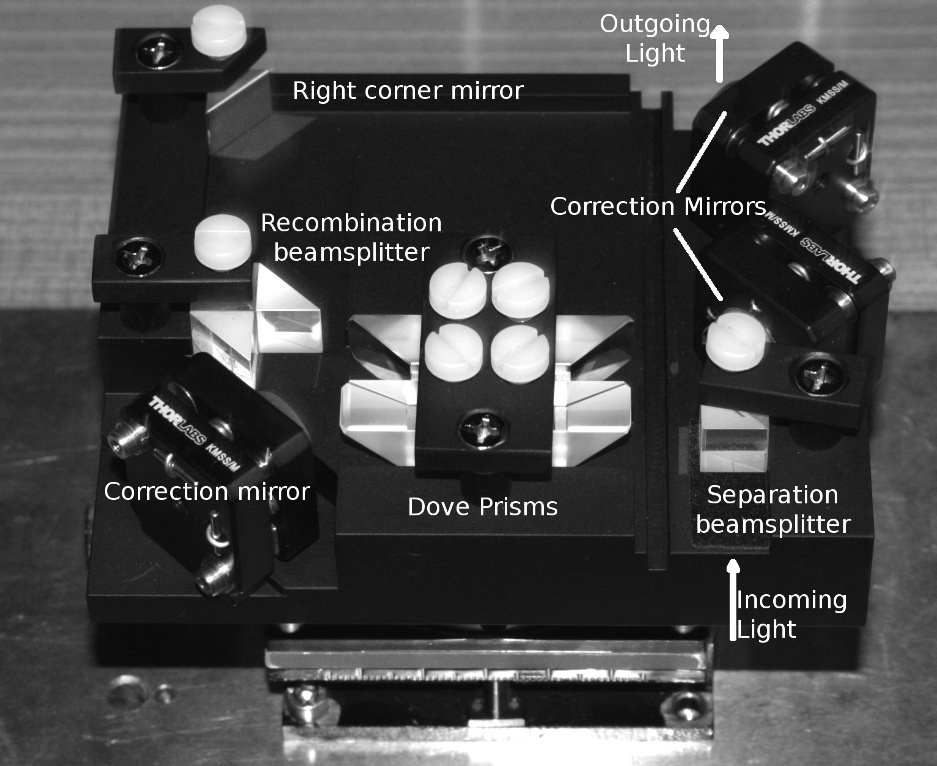}} 
  \hspace{2em}
   \subfloat[First Light image]{\label{fig:firstlight} \includegraphics[width = 0.36\textwidth]{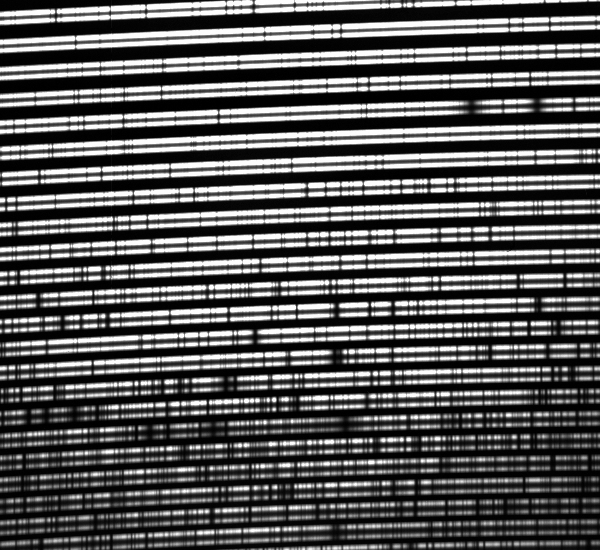}}
  \caption{ \emph{Left} \desspot prototype before insertion in front of the Coud\'e spectrograph. \emph{Right} First Light image taken for Capella under $0^\circ$ and $180^\circ$ orientation. The two orientations present the same line pattern in each order.}
  \label{fig:desspot_tls}
\end{figure}

A prototype of \desspot was implemented on the TLS-spectrograph. The only available location to insert \desspot in front of the slit was instead of the position of the iodine cell. 
The available space at this location is only 120 mm in the optical direction. Two fold mirrors were added in order to keep the optical axis, see Fig \ref{fig:desspot_tls}.
The insertion of \desspot in the light path resulted in a shift of the focus of 4 mm, which was easily compensated with the telescope. 

\subsubsection{Observations and results}

The observation targets were mostly super giants and spectroscopic binaries. Their signal is equivalent to the signal of a single star. In addition, their separation being larger than a star diameter, and their individual velocity being higher than the rotation velocity of a star without causing line broadening, the amplitude of their signal is bigger and easier to detect. They are observed in order to verify the method and to validate the instrumentation. We present here the results for Capella. \\
Capella is a system whose components are both giant stars, with similar magnitude and with moderate rotational velocities (4.5 $km s^{-1}$ and 35 $km s^{-1}$ for the A and B components \cite{Weber}). The orbit is well defined \cite{Torres} and the position angle of the system, i.e. the position angle of the ascending node $\Omega$, was determined with high accuracy.

Capella was observed in four runs spread over two nights. We used the natural field rotation of the Coud\'e modus to probe the stars under different orientation. The projected position angle of the slit is given by \begin{math} PA_{slit} = 151^\circ + \delta -\tau \end{math}, where $\delta$ is the stellar declination and $\tau$ its hour angle. We spanned the observations three hours apart each night till the slit position angle had changed by $90^\circ$. 
The probed orientations are resumed in Table \ref{tab:orient}. The maximum signal is expected when the slit is perpendicular to the system's position angle, which was almost achieved during the runs 2 and 4. 

\begin{table}
 \centering
  \begin{tabular}{p{3cm}|c|c|c|c|c}
    & Capella & run 1 & run 2 & run 3 & run 4 \\ \hline
 Date HJD (-245 000) & - & 5870.456497 & 5870.677959 & 5871.453430 & 5871.643367 \vspace{0.5em} \\ 
 Position angle & $ 40.421^\circ$ & $236.154^\circ$ & $156.307^\circ$ &  $236.225^\circ$ & $167.659^\circ$ \\
  \hline
  \end{tabular}
  \caption{The probed angles during the four observation runs, and the position angle of the Capella system.}
  \label{tab:orient}
\end{table}

A first glance on the images shows that for both orientations line position and intensity are similar, see Fig \ref{fig:firstlight}. The preparation of the images, e.g flatflield and scatter light corrections,  is done with IRAF. The intensity spectra are extracted by collapsing the orders along the slit direction, then Marsh algorithm\cite{Marsh} is applied for a correct estimation of the errors. Finally the intensity spectra are normalised to unity. The signal to noise ratio in these spectra is around 200. For the extraction of the position spectra, we wrote a program package which takes particular care of the correction of pixellation effects caused by the orders curvatures on the detector. 
 
\begin{figure}
 \centering
  \includegraphics[width = 0.9 \textwidth]{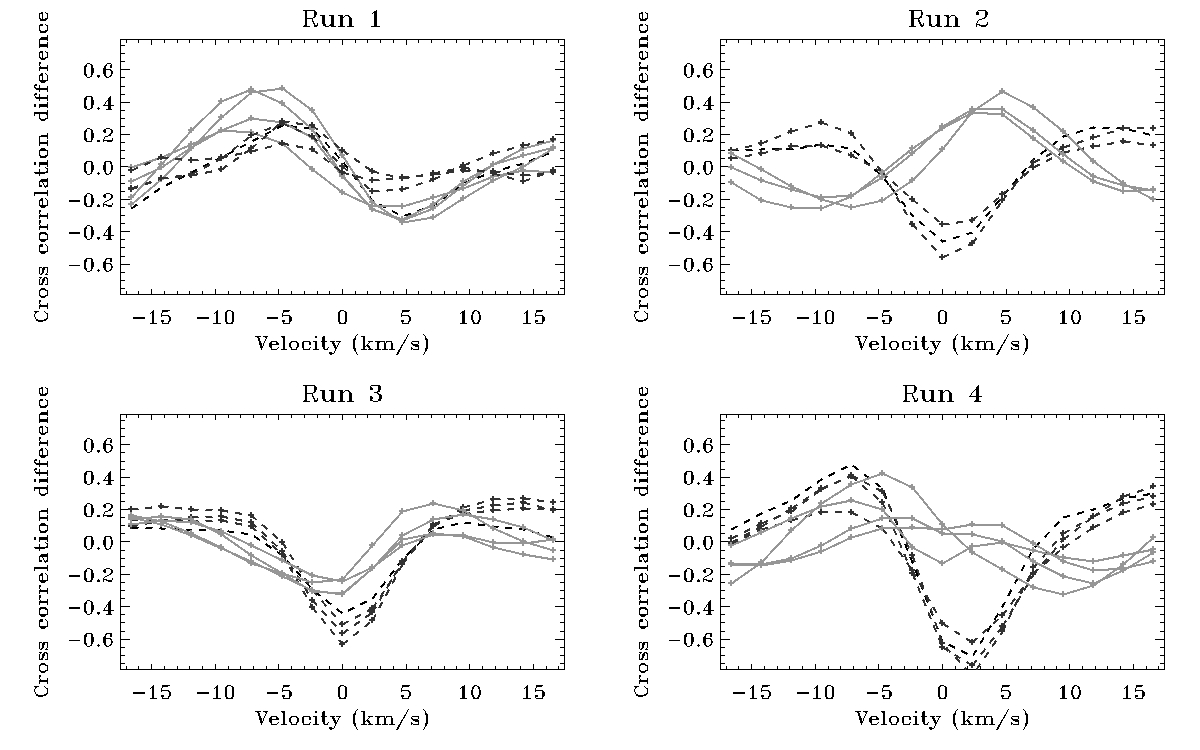}
  \caption{In black dotted lines, the cross-correlation function, for the orientation difference, for the telluric lines. In grey straight lines the one for the stellar lines. On the runs 1 and 3, there is a clear trend between the atmospheric lines and the stellar lines, while on the runs 2 and 4 the stellar lines have a significantly divengent cross-correlation function than the telluric lines.}
\label{fig:capesign}
\end{figure}

The signal of the line tilt is a shift of around 1-5 \% of a pixel in the position spectrum. Detecting it requires that this signal is measured with relatively low noise levels. This can not be achieved with a single line. Thus we proceed to the cross-correlation analysis between the intensity and the position spectrum. The cross-correlation functions of each orientation are then subtracted, i.e. the function from the $0^\circ$ orientation is subtracted from the function of the $180^\circ$ orientation. This differential function is to remain constant if no signal is expected. This procedure is repeated separately for the telluric and stellar lines. As illustrated in Fig \ref{fig:capesign} the differential cross-correlation function for the telluric lines is approximatively constant during one observation run. On the first and third run, the function for the stellar lines follows smoothly the trend of the function for the telluric lines. On the contrary, there are clear divergences between both functions in run 2 and 4. This coincides with the slit position with expected maximal signal. \\
We then determine the orbital position angle of Capella. The data is fitted with a sine curve of amplitude unity corresponding to the $sin \phi$ dependency of the line tilt. The least square solution for the phase yields $37.15^\circ \pm 5.25^\circ$ which is in agreement with the current value\cite{Torres} for the position angle: $40.421^\circ \pm 0.064^\circ$. This validates the instrumentation and the method for binary stars. With a dedicated spectrograph, we expect to reach the precision to determine the stellar spin position angle.

\section{THE DESSPOT-SPECTROGRAPH}

\subsection{Requirements}

Both \desspot instruments shall be tested on the 1.2-m telescope of the Hamburg Observatory. The Oskar-L\"uhning Telescope, hereafter the OLT,  has a Cassegrain output. The instruments shall be attached to the output, thus the spectrograph has to combine lightweight and compactness. 
We decide to use standard components whenever possible to construct the DeSSpOt-spectrograph, which reduces the cost to an acceptably level for a prototype. 

The DeSSpOt-spectrograph, hereafter DeSSpOtS, shall reach a resolution high enough to resolve the intrinsic stellar line width.

Our main targets are red super giants whose rotational velocities span over less than 1.5 kms$^{-1}$ to 35 kms$^{-1}$. In practice, stars with excessive macro-turbulences are discarded since it strongly affects the line widths. 
Therefore the needed resolution is evaluated around 50 000. The resolution of the spectrograph is given by 
\begin{equation}
 R = \frac{2 L \sin\theta_B}{ d_{slit}}  F/\#
\end{equation}
where L is the used length of the grating, $\theta_B$ the blaze angle, $d_{slit}$ the slit width and F/\# the incoming aperture ratio of the telescope. With the compactness limitation and the given aperture ratio of the telescope, the only current option we have, to reach the needed resolution, is to scale down the slit width to 22.8 $\mu$m. The spectrograph would have a limiting magnitude of around 4, which still includes most of the targeted red super giants.
In addition, the wavelength range spans from 460 nm to 700 nm, those limits are given by DeSSpOt. The spectrograph is set for the red part of this range. 
Finally, \desspots being a multi-channel spectrograph, the orders have to be separated accordingly. A minimal separation of 10 pixel between two channels on the detector is required to avoid the order overlap and enable the background determination. Moreover, we want to keep open the possibility of extending the number of channels in DeSSpOt. 

\subsection{Optical Design of \desspots}

\begin{figure}[h]
 \centering
 \includegraphics[width=0.75\textwidth]{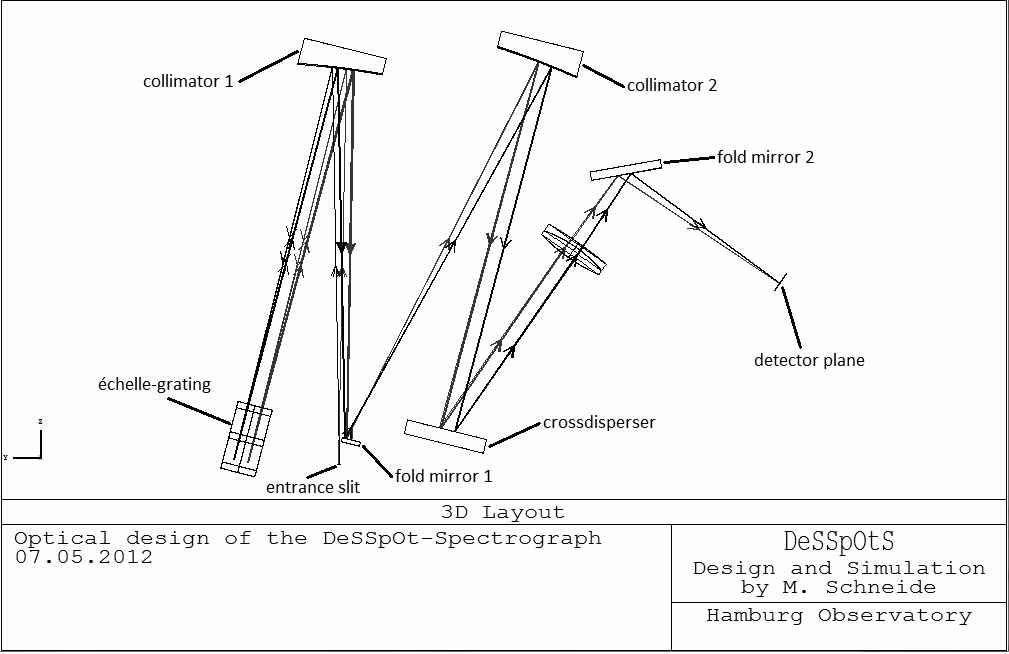}
 \caption[DeSSpOtS - Zemax]{Optical layout of the DeSSpOt-spectrograph. The simulation is done with Zemax and shows the basic concept for the spectroscopic multi-channel application.}
 \label{fig:desspotszmx}
\end{figure}

\desspots layout, as seen on Fig \ref{fig:desspotszmx}, is based on the proven white-pupil design which allows to reduce effectively the stray light in the system. The outgoing beam of the OLT is focused on the slit. Since the slit width has to be at most 22.8 $\mu$m in order to fulfil the requirements, we choose a fixed slit's width of 20 $\mu$m which matches a field of view of 0.27 arcsec in the sky. The light is collimated by an off-axis parabolic mirror with an effective focal length of 279.1 mm. The OLT output beam has an F/13 aperture ratio, which leads to a collimated beam diameter of 21.47 mm. \\
The beam is then reflected by the \'echelle-grating. It is a standard size aluminium coated grating with 52.67 lines/mm and a blaze angle $\theta_B$ of $63.5^\circ$. The ruled area has a length L of 49 mm which matches the beam projected diameter. The light is then sent back toward the first collimator. After passing collimator 1 the light is mirrored by a D-shaped fold mirror and collimated by the collimator 2, which is similar to the first one. \\
The cross-disperser is a reflection grating with 588 lines/mm chosen as to reach the required large order separation. 

Finally the beam is focused by an achromatic lens doublet with a focal length of 200 mm to image the \'{e}chelle-spectrum on the detector. We use an interline CCD from Apogee type ALTA U4000 with a 2k x 2k  chip and a pixel size of $7.4 \mu m$. By taking the optimal size of 2 pixel per spectral element, the field of view of one pixel is  0.135 arcsec/pix. 

Furthermore the length of the order at 600 nm is displayed on 1914 pixel whereas the order at 650 nm is displayed on 2074 pixel. Hence the length of the spectrum in the observed wavelength range fits well to the detector.

The complete set-up, including the telescope, \desspot and its spectrograph is simulated with Zemax, a ray-tracing program, to investigate the performance of the spectrograph.

\begin{figure}
 \centering
 \subfloat[Input of DeSSpOtS.]{\label{fig:slitplanezmx}\includegraphics[width=0.49\textwidth]{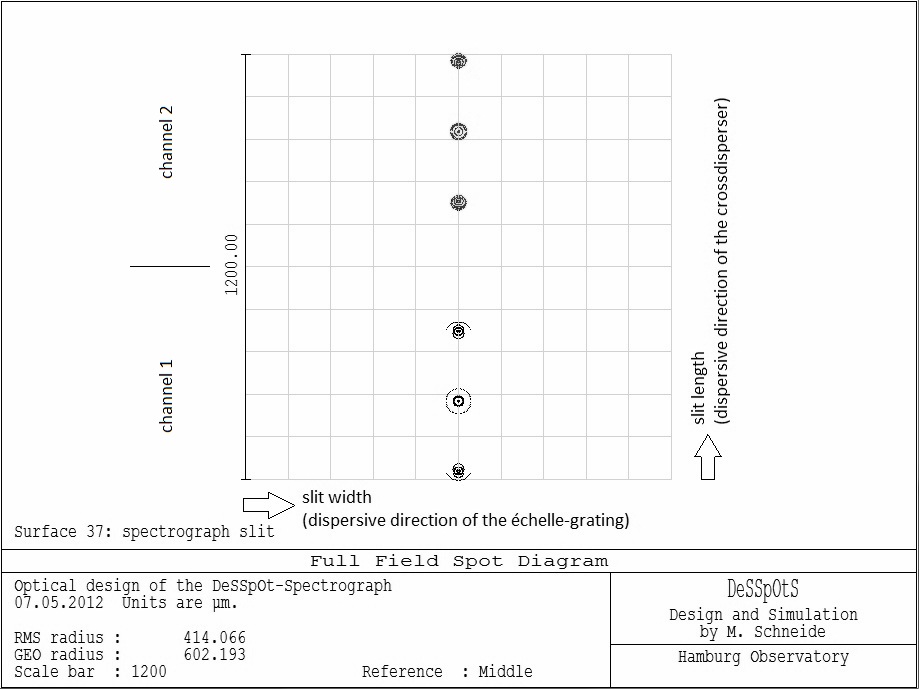}}
 ~
 \subfloat[Output of DeSSpOtS.]{\label{fig:imageplanezmx}\includegraphics[width=0.49\textwidth]{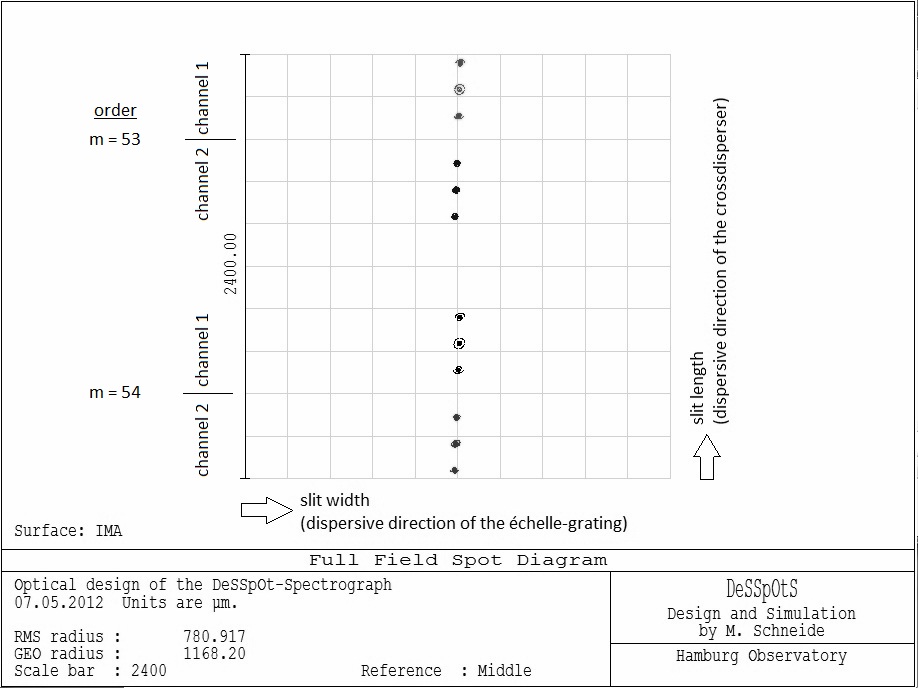}}
 \caption{\emph{Left:} Each channel represents one orientation of DeSSpOt with a 3 arcsec large star used as input. This gives a width of 400 $\mu$m for every channel. The separation between the two channels, as dictated by \desspot, is 760 $\mu$m. \emph{Right} Spot diagram of the output of the spectrograph for the orders 54 at 629.1 nm and order 53 at 641.18 nm. The orders are separated by 202 pixel. Each orientation is 41 pixel large.}
\label{fig:ioplanezmx}
\end{figure}

The following simulation example is done for the Littrow case, i.e. for the order 54 at 629.1 nm and the order 53 at 641.18 nm. The simulated spot diagram of the input focal plane at the entrance of the slit is depicted in Fig \ref{fig:slitplanezmx}. Each of the two channels consists of three spots where the central spot represents the optical axis of the OLT. The two outer spots delimit the width of a channel when imaging a 3 arcsec large star. The latter value corresponds to the average seeing conditions in Hamburg. The total width of an input channel is then estimated to 400 $\mu$m and the separation between the central spots of the channels to 760 $\mu$m. \\
The spot diagram of the output on the detector plane is depicted in Fig \ref{fig:imageplanezmx}. The width of a channel is imaged on 300 $\mu$m (41 pixel). The order separation is 1500 $\mu$m (202 pixel). As a result, ten complete orders are displayed on the detector. The channels within one order are separated by 36 pixel which is large enough for accurate background determination. In addition, there is enough space left for observations under inferior seeing conditions. 

\begin{figure}
 \centering
 \includegraphics[width=0.6\textwidth]{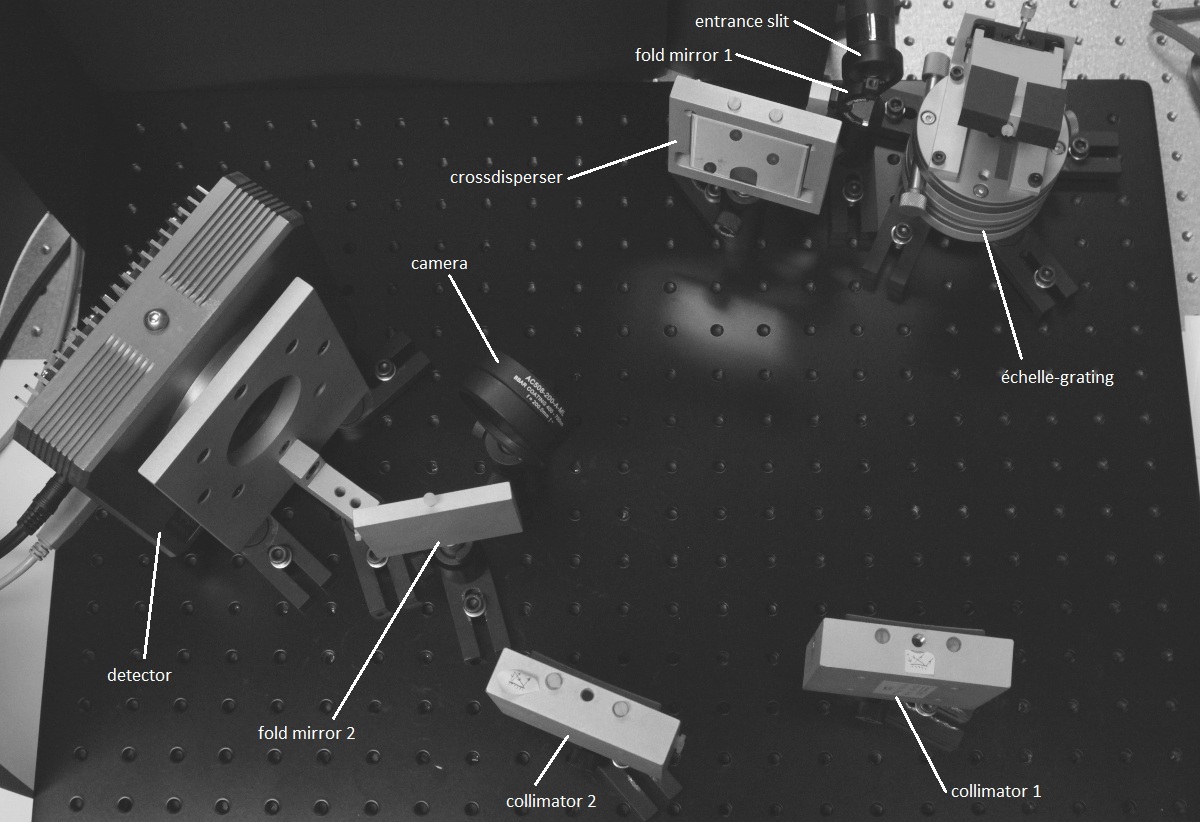}
 \caption[DeSSpOtS - Laboratory]{Current assembly of the DeSSpOt-spectrograph in the laboratory.}
 \label{fig:desspotslab}
\end{figure}

The whole DeSSpOt-spectrograph is constructed on an ultra light optical breadboard with a size of 450 x 600 $mm^2$ and a weight of about 8 kg. As shown in Fig \ref{fig:desspotslab} most of the components in the spectrograph are assembled with an optical post holder system but the \'{e}chelle-grating and the detector. The \'{e}chelle-grating is mounted on a rotary stage to be able to control the tiny tilt of the grating along the upright axis and on a goniometer that allows us to tilt the grating around the horizontal axis to adjust the blaze angle. The detector holder as well as some parts of the other holders are custom-made parts because there are no suitable standard components available. Altogether, the spectrograph will weight around 20 kg.

\subsection{First test results}

\begin{figure}
  \centering
  \subfloat[Imaging quality of the spectrograph]{\label{fig:neon_flat_slit}\includegraphics[width=0.42\textwidth]{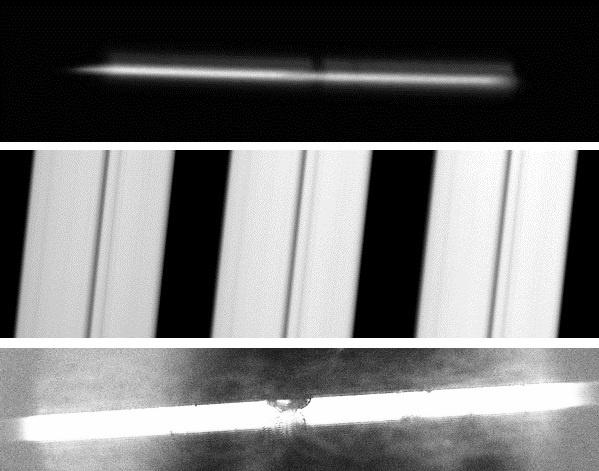}}
  \hspace{2em}
  \subfloat[Neon emission line profile at 653.288 nm.]{\label{fig:neonlineplot} \includegraphics[width=0.45\textwidth]{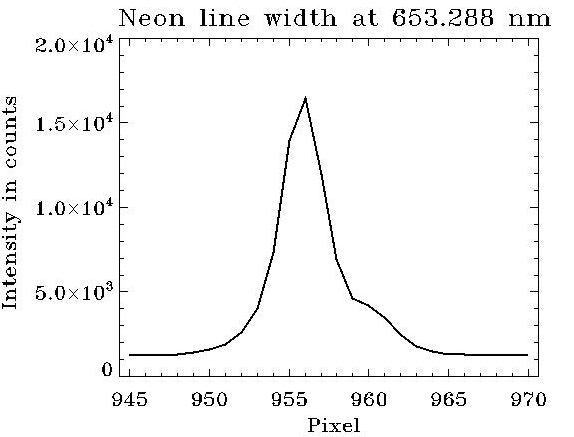}}
  \caption{\emph{Top left:} This image shows the neon emission line at 653.288 nm on the detector. \emph{Middle left:} A cut-out of a flat field spectrum with obscurations caused by the entrance slit. \emph{Bottom left:} An image of the entrance slit showing the jagged edges. \emph{Right:} This image shows the profile plot of the neon emission line.}
  \label{fig:calibs}
\end{figure}

The spectrograph is currently tested in the laboratory without \desspot. In this chapter the current imaging quality is discussed on the basis of the first test results. \\
The spectrograph is calibrated with a neon lamp. As illustrated on top of the Fig \ref{fig:neon_flat_slit}, the neon emission line presents a certain blurring and distortion. This is identified as a consequence of the non-perfect adjustment of the off-axis parabolic mirrors. In addition, the middle of the neon line is obscured despite the homogeneous illumination of the entrance slit. This pattern is present in the entire spectrum, for whatever illumination,  as depicted in the flat field cut-out in the middle of Fig \ref{fig:neon_flat_slit}. A magnified image of the slit at the bottom of Fig \ref{fig:neon_flat_slit} shows that it has unclean edges with a large jag in the middle leading to a strong diminution of the slit width.  \\
The profile of the neon line, presented in Fig \ref{fig:calibs}, is much larger than the 2 pixel requirement for one spectral element. This broadening, as well as the bump on the right side of the line around pixel number 960, is mostly caused by astigmatism.

\begin{figure}
  \centering
  \includegraphics[width=0.98\textwidth]{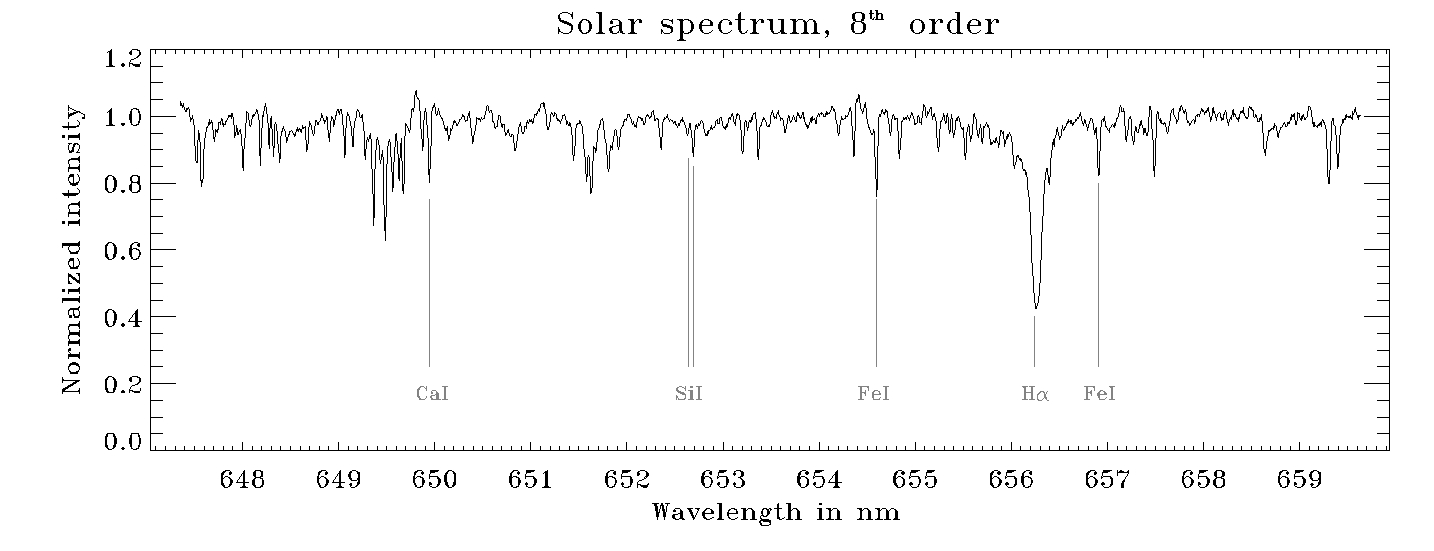}
  \caption[Solar spectrum]{Solar spectrum in order $52$ of DeSSpOtS showing some resolved and identified absorption lines. Order $52$ of 
the \'{e}chelle-grating equates to the $8^{th}$ of $10$ orders on the 
detector.}
  \label{fig:solarspec}
\end{figure}

We also captured a spectrum of the sun by transporting the light to the laboratory by a fibre. The signal to noise achieved in the spectrum peaks at 150. The $H_\alpha$ line is imaged in the order 52, corresponding to the $8^{th}$ order on the detector, as seen in Fig \ref{fig:solarspec}. In addition, we identify some FeI, SiI and CaI lines. In its current configuration, the DeSSpOt-spectrograph is already able to resolve several useful absorption lines for stellar spin determination.
An analysis of the spatial component of the spectrum will show if this configuration of DeSSpOtS is sufficient. Therefore further tests with direct feeding of the spectrograph are required.

\section*{CONCLUSION}

We have successfully implemented the differential image rotator \desspot to the Coud\'e-spectrograph at the Th\"uringer Landessternwarte. With our observations of Capella, we found a orbital position angle of $37.15^\circ$ in agreement with the value found in the literature. As such the method for the determination of stellar position angle is verified. 
We also presented the primary design of the DeSSpOt-spectrograph,  dedicated to \desspot for applications at Hamburg Observatory.

\acknowledgments

We would like to thank here E. G\"unther, H. Lehmann and A. Hatzes of the Th\"uringer Landessterwarte Tautenburg for giving us the opportunity to implement \desspot on their spectrograph. In addition, we gratefully acknowledge the rapid and accurate work of J. Winckler and T. L\"owinger of their  mechanical workshop which got involved in this project. This project was partly funded by the German Federal Ministry for Research and Technology BMBF grant 05A08BU2.

\bibliography{lesagespie}

\begin{thebibliography}{10}

\bibitem{Beckers}
{Beckers}, J.~M., ``{Differential speckle interferometry.},'' {\em Optica
  Acta}~{\bf 29},  361--362 (Apr. 1982).

\bibitem{Lagarde}
{Lagarde}, S., {Sanchez}, L.~J., and {Petrov}, R.~G., ``{Sub-resolution Limit
  Spatio-Spectral Information Using Differential Speckle Interferometry},'' in
  [{\em IAU Colloq. 149: Tridimensional Optical Spectroscopic Methods in
  Astrophysics}{\nolinebreak\hspace{0.1em}]},  {G.~Comte \& M.~Marcelin}, ed.,
  {\em Astronomical Society of the Pacific Conference Series} {\bf 71},  360
  (1995).

\bibitem{Monnier}
{Monnier}, J.~D., {Zhao}, M., {Pedretti}, E., {Thureau}, N., {Ireland}, M.,
  {Muirhead}, P., {Berger}, J.-P., {Millan-Gabet}, R., {Van Belle}, G., {ten
  Brummelaar}, T., {McAlister}, H., {Ridgway}, S., {Turner}, N., {Sturmann},
  L., {Sturmann}, J., and {Berger}, D., ``{Imaging the Surface of Altair},''
  {\em Science}~{\bf 317},  342-- (July 2007).

\bibitem{deSouza}
{Domiciano de Souza}, A., {Kervella}, P., {Jankov}, S., {Abe}, L., {Vakili},
  F., {di Folco}, E., and {Paresce}, F., ``{The spinning-top Be star Achernar
  from VLTI-VINCI},'' {\em A\&A}~{\bf 407},  L47--L50 (Aug. 2003).

\bibitem{Peterson}
{Peterson}, D.~M., {Hummel}, C.~A., {Pauls}, T.~A., {Armstrong}, J.~T.,
  {Benson}, J.~A., {Gilbreath}, G.~C., {Hindsley}, R.~B., {Hutter}, D.~J.,
  {Johnston}, K.~J., {Mozurkewich}, D., and {Schmitt}, H.~R., ``{Vega is a
  rapidly rotating star},'' {\em Nature}~{\bf 440},  896--899 (Apr. 2006).

\bibitem{Zhao}
{Zhao}, M., {Monnier}, J.~D., {Pedretti}, E., {Thureau}, N., {M{\'e}rand}, A.,
  {ten Brummelaar}, T., {McAlister}, H., {Ridgway}, S.~T., {Turner}, N.,
  {Sturmann}, J., {Sturmann}, L., {Goldfinger}, P.~J., and {Farrington}, C.,
  ``{Imaging and Modeling Rapidly Rotating Stars: {$\alpha$} Cephei and
  {$\alpha$} Ophiuchi},'' {\em ApJ}~{\bf 701},  209--224 (Aug. 2009).

\bibitem{Whelan}
{Whelan}, E. and {Garcia}, P., ``{Spectro-astrometry: The Method, its
  Limitations, and Applications},'' in [{\em Jets from Young Stars
  II}{\nolinebreak\hspace{0.1em}]},  {F.~Bacciotti, L.~Testi, \& E.~Whelan},
  ed., {\em Lecture Notes in Physics, Berlin Springer Verlag} {\bf 742},  123
  (2008).

\bibitem{Bailey}
{Bailey}, J., ``{Detection of pre-main sequence binaries using
  spectro-astrometry},'' {\em MNRAS}~{\bf 301},  161--167 (1998).

\bibitem{Voigt}
{Voigt}, B. and {Wiedemann}, G., ``{Surface spots on cool giants probed by
  spectro-astrometry},'' {\em American Institute of Physics Conference Series}
  {\bf 1094},  896--899 (Feb. 2009).

\bibitem{Denisov}
{Denisov}, N.~A. and {Koroleva}, T.~V., ``{Design of mirror image rotator
  systems},'' {\em Society of Photo-Optical Instrumentation Engineers (SPIE)
  Conference Series} {\bf 3055},  297--300 (Feb. 1997).

\bibitem{Sarel}
{Sar-El}, H.~Z., ``{Revised Dove prism formulas},'' {\em Applied Optics}~{\bf
  30},  375--376 (Feb. 1991).

\bibitem{Strojnik}
{Moreno}, I., {Paez}, G., and {Strojnik}, M., ``{Dove prism with increased
  throughput for implementation in a rotational-shearing interferometer},''
  {\em Applied Optics}~{\bf 42},  4514--4521 (Aug. 2003).

\bibitem{roe}
{Roe}, H.~G., ``{Implications of Atmospheric Differential Refraction for
  Adaptive Optics Observations},'' {\em Astronomical Society of the
  Pacific}~{\bf 114},  450--461 (Apr. 2002).

\bibitem{Weber}
{Weber}, M. and {Strassmeier}, K.~G., ``{The spectroscopic orbit of Capella
  revisited},'' {\em A \& A}~{\bf 531},  A89 (July 2011).

\bibitem{Torres}
{Torres}, G., {Claret}, A., and {Young}, P.~A., ``{Binary Orbit, Physical
  Properties, and Evolutionary State of Capella ({$\alpha$} Aurigae)},'' {\em
  apj}~{\bf 700},  1349--1381 (Aug. 2009).

\bibitem{Marsh}
{Marsh}, T.~R., ``{The extraction of highly distorted spectra},'' {\em
  Astronomical Society of the Pacific}~{\bf 101},  1032--1037 (Nov. 1989).

\end{thebibliography}
\bibliographystyle{spiebib}

\end{document}